\documentclass[hyper]{JHEP} 

\usepackage{epsfig}

\newcommand\fverb{\setbox\pippobox=\hbox\bgroup\verb}

\newcommand\fverbdo{\egroup\medskip\noindent%

            \fbox{\unhbox\pippobox}\ }

\newcommand\fverbit{\egroup\item[\fbox{\unhbox\pippobox}]}

\newbox\pippobox

\title{Canonical Description of T-duality with
NSNS Background}
\author{J. Kluso\v{n}\\
Department of
Theoretical Physics and Astrophysics\\
Faculty of Science, Masaryk University\\
Kotl\'{a}\v{r}sk\'{a} 2, 611 37, Brno\\
Czech Republic\\
E-mail: \email{klu@physics.muni.cz}} \preprint{}

 \abstract{We study T-duality with non-zero components of  NSNS two form field
    along directions we dualize with the help of canonical formalism. As a result of this procedure we determine generalized Buscher's rules. We also
 apply the same procedure to  the case of non-relativistic string.}

\def\hM{\hat{M}}
\def\hN{\hat{N}}

\def\tp{\tilde{p}}

\def\hG{\hat{G}}

\def\bH{\mathbf{H}}

\def\bA{\mathbf{A}}

\def\bV{\mathbf{V}}

\def\ty{\tilde{y}}

\def\tlambda{\tilde{\lambda}}
\def\bB{\mathbf{B}}

\def\tx{\tilde{x}}

\def\be{\begin{equation}}

\def\ee{\end{equation}}

\def\bea{\begin{eqnarray}}

\def\eea{\end{eqnarray}}

\def\bX{\mathbf{X}}
\def\bY{\mathbf{Y}}

\def\hN{\hat{N}}

\def\mH{\mathcal{H}}

\def\tG{\tilde{G}}

\newcommand{\tX}{\tilde{X}}

\newcommand{\hh}{\hat{h}}

\def \bA{\mathbf{A}}

\newcommand{\ba}{\mathbf{a}}

\newcommand{\mL}{\mathcal{L}}

\def\pb #1{\left\{#1\right\}}

\begin{document}
\section{Introduction and Summary}
One of the most important properties of string theory is T-duality,
for earlier review, see \cite{Giveon:1994fu}. This is duality of
string theory whose most powerful description is given in terms of
Buscher's rules \cite{Buscher:1987sk,Buscher:1987qj} of the
transformations of the background fields under T-duality. More
explicitly, we start with string sigma model on the background where
the background metric possesses one isometry, at least. Then we
gauge this isometry so that this is now local symmetry on the string
world-sheet when we introduce corresponding covariant derivative and
two dimensional gauge field which are non-propagating. In order to
ensure that these fields are not dynamical we add to the action term
that ensures that the field strength of this gauge field is zero. As
the next step we fix the gauge when we take the world-sheet mode
that parameterizes direction with gauged isometry, to be zero. Then
we can solve the flatness of the gauge field with introducing new
scalar mode that parameterizes the string propagating along dual
coordinate where now the background fields are related to the
original ones through Buscher's rules.

While this procedure is well established in case of the T-duality
along one direction  situation when we have several isometry
directions with non-trivial NSNS two form is more complicated. In
our previous paper \cite{Kluson:2019xuo} we encountered this problem
when we analyzed T-duality of non-relativistic string in torsional
Newton-Cartan background. Since such a string can be defined as
T-dual of the relativistic string in the background with light-like
isometry
\cite{Harmark:2017rpg,Kluson:2018egd,Harmark:2018cdl,Harmark:2019upf}
in order to study its T-duality properties we had to analyze
T-duality of relativistic string along two directions. Even if such
a procedure seems to be straightforward it turned out to be rather
involved and we were not able to determine transformations rules in
the general case of non-zero components of NSNS two form along
directions we dualize. Surprisingly we were not able to find
corresponding transformations rules in the literature and the goal
of this paper is to find them.

Explicitly, we would like to study general string theory action in the background with non-trivial metric and NSNS two form with isometry along several directions. In principle we could start with the Lagrangian formulation and gauge corresponding isometry directions and then proceed as in the case of single isometry direction. However now the situation is much more intricate due to the presence of non-trivial NSNS two form that now appears in corresponding equations of motions. It is important to stress that expressions with NSNS two form are multiplied with world-sheet antisymmetric symbol while expressions with target space metric are multiplied with world-sheet metric
so that it is very difficult to solve them. However there is an alternative
way how to study T-duality which is  canonical approach to T-duality
 \cite{Alvarez:1994wj,Alvarez:1994dn}. This procedure is based on the
Hamiltonian form of string in the background that possesses an
isometry.  Then we perform canonical transformations of the
coordinate that labels this isometric direction and we derive new
Hamiltonian for dual theory. Finally we perform inverse Legendre
transformations to T-dual Lagrangian and we determine dual
background fields whose forms agree with Buscher's rules. We use
this procedure to the case of the relativistic string in the
background with non-zero NSNS two form. We determine T-dual
Hamiltonian which is straightforward procedure. On the other hand in
order to determine corresponding transformation rules for the
background field we have to find T-dual Lagrangian. It turns out
that this is non-trivial procedure that deserves careful treatment.
After performing this analysis we obtain generalized Buscher's rules
that also include background NSNS two form. As far as we know these
transformations were not determined before in the full generality.

As the next goal of this work is to analyze T-duality of
non-relativistic string in torsional Newton-Cartan background and
with non-trivial NSNS two form. This section is generalization of
the analysis performed recently in \cite{Kluson:2019xuo}. We show
that generally under T-duality non-relativistic string maps to the
relativistic one with specific form of the background fields and we
also analyze conditions that determine that  non-relativistic string
maps to non-relativistic string again. We show that these conditions
are the same as ones that were found in \cite{Kluson:2019xuo}.

Let us outline our result and suggest possible extension of this work. We study T-duality along several directions with non-zero NSNS two form in the framework of canonical formalism. We obtain corresponding generalized Buscher's rules which as far as we know, were determined for the first time in this generality. Then we analyze T-duality of non-relativistic string in torsional Newton-Cartan background with non-zero NSNS two form and we determined conditions under such a string maps to non-relativistic string. We mean that the analysis presented in this paper could be extended in several directions. In particular, it would be nice to study whether T-dual Hamiltonian found in this paper could be useful for the symmetric formulation of the string in the Double string theory. It would be also nice whether similar analysis could be performed in case of the non-relativistic string. We hope to return to these problems in future.

The structure of this paper is as follows. In the next section
(\ref{second}) we analyze T-duality of relativistic string with the
help of canonical formalism. Then in section (\ref{third}) we
present the same analysis in case of the non-relativistic string in
torsional Newton-Cartan background.

\section{T-duality with NSNS two form in Canonical Formalism}\label{second}
T-duality with non-trivial NSNS two form along directions we dualize
is remarkably complex if we consider general bosonic string. The
reason is that the term which is proportional to the embedding of
the target space metric is multiplied by word-sheet metric while
term which is proportional to NSNS two form is multiplied by
world-sheet antisymmetric tensor. Then it is very difficult to solve
equations of motion for auxiliary world-sheet gauge fields. However
the situation simplifies considerably when we analyze T-duality with
the help of the canonical formalism
   \cite{Alvarez:1994wj,Alvarez:1994dn}. To do this we have to find Hamiltonian for
 bosonic string with the action
    \begin{equation}
   S=-T\int d^2\sigma \sqrt{-\gamma}\gamma^{\alpha\beta}G_{MN}
 \partial_\alpha x^M
 \partial_\beta x^N- \frac{T}{2}\int d^2\sigma
 \epsilon^{\alpha\beta}B_{MN}\partial_\alpha x^M\partial_\beta x^m\ ,
 \end{equation}
where $T$ is string tensor, $\gamma_{\alpha\beta}$ is world-sheet metric, $x^M,M=0,\dots,25$ parameterize embedding of the string in the target space background
with the metric $G_{MN}$ and NSNS two form $B_{MN}$. Further, world-sheet is parameterized with $\sigma^\alpha,\alpha=0,1, \sigma^0=\tau \ , \sigma^1=\sigma$ and $\epsilon^{\alpha\beta}=-\epsilon^{\beta\alpha}$ is antisymmetric tensor.

Since we are going to analyze T-duality with the help of the
canonical formalism we have to introduce Hamiltonian for bosonic
string. The result is well known and we write the result as
\begin{equation}
H=\int d\sigma (N^\tau\mH_\tau+N^\sigma \mH_\sigma) \ ,
\end{equation}
where
\begin{eqnarray}
& &\mH_\tau=\Pi_M G^{MN}\Pi_N+T^2 G_{MN}\partial_\sigma x^M\partial_\sigma X^N \ ,
\quad
\Pi_M=p_M+TB_{MN}\partial_\sigma x^N \ ,
\nonumber \\
& &\mH_\sigma=p_M\partial_\sigma x^N \ ,
\end{eqnarray}
where $N^\tau,N^\sigma $ are Lagrange multipliers corresponding to
the first class constraints $\mH_\tau\approx 0 \ , \mH_\sigma\approx
0$.

Now we are ready to proceed to the canonical description of T-duality when we select $2p$ coordinates $x^m, m,n=25-2p,\dots,25$ and the remaining ones $\mu,\nu=0,1,\dots,2p-1$ and dualize along $x^m$ directions. According to  \cite{Alvarez:1994wj,Alvarez:1994dn}
such a duality can be considered as canonical transformation. Explicitly, we introduce T-dual variables $\tx_m$ and corresponding conjugate momenta
$\tp^m$. Then the canonical transformations have the form
\begin{equation}\label{Tdualrul}
\tp^m=-T\partial_\sigma x^m  \ , \quad p_m=-T\partial_\sigma \tx^m
\end{equation}
and T-dual Hamiltonian arises when we replace original variables
$p_m,x^m$ in the Hamiltonian constraint with T-dual ones given
above. As a result we obtain following T-dual Hamiltonian constrain
in the form
\begin{eqnarray}\label{mHtau}
& &\mH_\tau^T=(k_\mu-B_{\mu m}\tp^m)G^{\mu\nu}(k_\nu-B_{\nu n}\tp^n)+2(k_\mu-B_{\mu m}\tp^m)G^{\mu n}
(-T\partial_\sigma \tx_n-B_{nm}\tp^m+TB_{n\mu}
\partial_\sigma x^\mu)+\nonumber \\
& &(-T\partial_\sigma \tx_m-B_{mk}\tp^k+TB_{m\mu}
\partial_\sigma x^\mu )G^{mn}
(-T\partial_\sigma \tx_n-B_{nl}\tp^l+TB_{n\nu}
\partial_\sigma x^\nu)+\nonumber \\
& &+T^2 \partial_\sigma x^\mu G_{\mu\nu}\partial_\sigma x^\nu-2T\tp^m G_{m\mu}
\partial_\sigma x^\mu+\tp^m G_{mn}\tp^n= \nonumber \\
& &=k_\mu G^{\mu\nu}k_\nu +2\tp^m B_{mM}G^{M\mu}k_\mu+
\tp^m\bH_{mn}\tp^n+2Tk_\mu G^{\mu n}\bV_n+
\nonumber \\
& &+T^2\bV_m G^{mn}\bV_n+2T\tp^k (B_{kM}G^{Mn}\bV_n-G_{m\mu}\partial_\sigma x^\mu)
+T^2 \partial_\sigma x^\mu G_{\mu\nu}\partial_\sigma x^\nu \ , \nonumber \\
\end{eqnarray}
where
\begin{equation}
k_\mu=p_\mu+TB_{\mu\nu
}\partial_\sigma x^\nu, \  \quad \bV_n=-\partial_\sigma \tx_n+
B_{n\mu}\partial_\sigma x^\mu \ ,
\end{equation}
and where we used the fact that
under T-duality (\ref{Tdualrul}) $\Pi_M$ transform as
\begin{eqnarray}
\Pi_\mu=k_\mu-B_{\mu m}\tp^m \ ,
\quad
\Pi_m=-T\partial_\sigma \tx_m-B_{mn}\tp^n+TB_{m\mu}
\partial_\sigma x^\mu \ . \nonumber \\
\end{eqnarray}
As the last important step we introduced matric $\bH_{mn}$ defined as
\begin{equation}
\bH_{mn}=G_{mn}-B_{mM}G^{MN}B_{Nm} \ .
\end{equation}
In order to determine transformation rules for metric and NSNS two form
field under these T-duality transformations we have to find Lagrangian
for T-dual string. With the help of the Hamiltonian constraint
(\ref{mHtau}) we obtain following
 equation of motion
\begin{eqnarray}\label{partialtauxm}
& &\partial_\tau \tx_m=\pb{\tx_m,H}=
2N^\tau B_{mM}G^{M\mu}k_\mu+2N^\tau \bH_{mn}\tp^n+\nonumber \\
&&+
2TN^\tau (B_{mM}G^{Mn}\bV_n-G_{m\mu}\partial_\sigma x^\mu)+N^\sigma
\partial_\sigma \tx_m \ , \nonumber \\
& &\partial_\tau x^\mu=\pb{x^\mu,H}=
2N^\tau G^{\mu\nu}k_\nu-2N^\tau G^{\mu M}B_{Mm}\tp^m+
2N^\tau T G^{\mu n}\bV_n \ . \nonumber \\
\end{eqnarray}
From the second equation we express $k_\mu$ as
\begin{equation} \label{kmu}
k_\mu=\frac{1}{2N^\tau}\hG_{\mu\nu}(X^\nu+2N^\tau G^{\nu M}B_{Mm}\tp^m-2N^\tau
T G^{\mu n}\bV_n) \ ,
\end{equation}
where
\begin{equation}
\hG_{\mu\nu}=G_{\mu\nu}-G_{\mu m}\hG^{mn}G_{n\nu} \ ,  \quad
\tX^\mu=\partial_\tau x^\mu-N^\sigma \partial_\sigma x^\mu \ , \quad
\tX_m=\partial_\tau \tx_m-N^\sigma\partial_\sigma \tx_m \ ,
\end{equation}
and where we introduced matrix $\hG^{mn}$ that is inverse to the matrix
$G_{mn}$
\begin{equation}
 \hG^{mn}G_{nk}=\delta^m_k \ .
 \end{equation}
 Further, matrix $\hG_{\mu\nu}$ has following important properties
\begin{equation}
\hG_{\mu\nu}G^{\nu\rho}=\delta_\mu^\rho \ , \quad
\hG_{\mu \nu}G^{\nu m}=-G_{\mu m}\hG^{mn} \ .
\end{equation}
Then inserting (\ref{kmu}) into the first equation in (\ref{partialtauxm}) we obtain
\begin{eqnarray}\label{tXm}
\tX_m=2N^\tau \hG_{mn}\tp^n+2N^\tau T B_{ml}\hG^{ln}\bV_n+B_{m\mu}X^\mu-B_{mn}\hG^{nk}G_{k\nu}X^\nu-2TN^\tau G_{m\mu}\partial_\sigma x^\mu \ ,
\nonumber \\
\end{eqnarray}
where we used following important properties
\begin{eqnarray}
& &B_{mM}G^{M\mu}\hG_{\mu\nu}G^{\nu N}B_{Nn}=B_{mM}G^{MN}B_{Nn}-B_{mk}
\hG^{kl}B_{ln} \ , \nonumber \\
& &-B_{mM}G^{M\mu}\hG_{\mu\nu}G^{\nu n}\bV_n=B_{ml}\hG^{ln}\bV_n-
B_{mM}G^{Mn}\bV_n \ , \nonumber \\
\end{eqnarray}
and where we defined
\begin{equation}
\hG_{mn}=G_{mn}-B_{mk}\hG^{kl}B_{ln} \ .
\end{equation}
Now it is easy to express $\tp^m$ from (\ref{tXm})
\begin{eqnarray}
\tp^m=\frac{1}{2N^\tau}\tG^{mn}(\tX_n-B_{n\mu}X^\mu+B_{nk}\hG^{kl}G_{l\nu}X^\nu+2TN^\tau
G_{m\mu}\partial_\sigma x^\mu-2N^\tau T                                  B_{ml}\hG^{ln}\bV_n) \ ,
\nonumber \\
\end{eqnarray}
where $\tG^{mn}$ is matrix inverse to $\hG_{mn}$ defined as
\begin{equation}
\tG^{mn}\hG_{nk}=\delta^m_k \ .
\end{equation}
An existence of the matric $\tG^{mn}$ is crucial consequence of the presence
of non-trivial components of NSNS two form $B_{mn}$. Clearly for $B_{mn}=0$, $\tG^{mn}$ reduces to $\hG^{mn}$.
%
%
%
Now we are ready to find corresponding Lagrangian density
\begin{eqnarray}\label{LTdual}
& &\mL=p_\mu \partial_\tau x^\mu+\tp^m\partial_\tau \tx_m-N^\tau \mH_\tau^T-
N^\sigma \mH^T_\sigma=
\nonumber \\
 && =\frac{1}{4N^\tau}[X^\mu G'_{\mu\nu}X^\nu+\tX_m G'^{mn}\tX_n+X^\mu G'^{ \ m}_\mu \tX_m+
\tX_n G'^n_{ \ \nu}X^\nu]-\nonumber \\
& &-T^2N^\tau[\partial_\sigma \tx_m G'^{mn}\partial_\sigma \tx_n+
\partial_\sigma x^\mu G'_{\mu\nu}\partial_\sigma x^\nu+\partial_\sigma \tx_m
G'^m_{ \ \nu}\partial_\sigma x^\nu
+\partial_\sigma x^\mu G'^{ \ n}_\mu\partial_\sigma \tx_n] \nonumber \\
& &-TB'_{\mu\nu}\partial_\tau x^\mu\partial_\sigma x^\nu-
TB'^{ \ n}_{\mu }\partial_\tau x^\mu\partial_\sigma \tx_n-
TB'^m_{ \ \nu}\partial_\tau \tx_m \partial_\sigma \tx^\nu
-TB'^{mn}\partial_\tau \tx_m\partial_\sigma \tx_n \equiv
\nonumber \\
& &
\equiv \frac{1}{4N^\tau}[g'_{\tau\tau}-2N^\sigma g'_{\tau\sigma}+(N^\sigma)^2g'_{\sigma\sigma}]-T^2 N^\tau g'_{\sigma\sigma}-Tb'_{\tau\sigma} \ , \nonumber \\
\end{eqnarray}
where we used an important property of the matrix $\tG^{mn}$
\begin{equation}
\tG^{mn}=\hG^{mn}+\hG^{mk}B_{kl}\tG^{lp}B_{pr}\hG^{rn}  \ , \quad \tG^{mn}B_{nk}\hG^{kl}=
\hG^{mn}B_{nk}\tG^{kl} \ .
\end{equation}
Finally the components of T-dual metric and NSNS two form have the form
\begin{eqnarray}\label{genBuschers}
& &G'_{\mu\nu}
=G_{\mu\nu}-G_{\mu m}\hG^{mn}G_{n\nu}
-(B_{\mu m}-
G_{\mu l}\hG^{lk}B_{km})
\tG^{mn}(B_{n\nu}-B_{nk'}\hG^{k'l'}G_{l'\nu}) \ , \nonumber \\
& &G'^{mn}=\tG^{mn} \ , \quad  G'^m_{ \ \mu}=-\tG^{mn}
(B_{n\nu}-B_{nk}\hG^{kl}G_{l\nu}) \ , \quad
G'^{ \ n}_\mu=
(B_{\mu n}-G_{\mu k}\hG^{kl}B_{lm})\tG^{mn} \ .
\nonumber \\
& &B'^{mn}=\tG^{mk}B_{kl}\hG^{ln} \  , \nonumber \\
& & B'_{\mu\nu}=
B_{\mu\nu}+ G_{\mu m}
\tG^{mn}B_{n\nu}-B_{\mu m}\tG^{mn}G_{n\nu}+B_{\mu m}\tG^{mn}B_{nl}\hG^{lk}B_{k\nu}+G_{\mu k}\hG^{kl}B_{lm}
\tG^{mn}G_{n\nu} \ , \nonumber \\
& &B'^{ \ n}_\mu=
[G_{\mu m}-B_{\mu k}\hG^{kl}B_{lm}
]\tG^{mn} \ , \quad B'^m_{ \ \mu}=
\tG^{mn} [G_{n\mu}-B_{nk}\hG^{kl}B_{l\mu}] \ . \nonumber \\
\end{eqnarray}
These are most general T-duality transformation rules in case
of non-zero components of NSNS two form along directions where T-duality
is performed. We see presence of two inverse metrics $\hG^{mn}$ and $\tG^{mn}$ where
$\tG^{mn}$ reduces into $\hG^{mn}$ in case of zero $B_{mn}$. In fact, in this case
the transformation rules reduce to transformation rules found in the previous paper
\cite{Kluson:2019xuo}. Finally $g'_{\alpha\beta},b'_{\alpha\beta}$ introduced on the last line in (\ref{LTdual}) are pullbacks of T-dual components of metric and NSNS two form
given in (\ref{genBuschers}) to the string's world-sheet. Finally, if we integrate out $N^\tau,N^\sigma$ we obtain
\begin{equation}
N^\sigma=\frac{g'_{\tau\sigma}}{g'_{\sigma\sigma}} \ , \quad
N^\tau=\frac{1}{2T}\frac{\sqrt{-\det g'_{\alpha\beta}}}{g'_{\sigma\sigma}} \ .
\end{equation}
Then inserting these results back to (\ref{LTdual}) we obtain Nambu-Goto form of the string action in T-dual background (\ref{genBuschers})
\begin{equation}
\mL=-T\sqrt{-\det g'_{\alpha\beta}}-Tb'_{\tau\sigma} \ .
\end{equation}
 In the next section we focus on T-duality with non-zero NSNS two
form in case of non-relativistic string.

\section{T-duality of Non-Relativistic String }\label{third}
In this section we will briefly discuss T-duality properties of non-relativistic string
in torsional Newton-Cartan background with the presence of non-zero NSNS two form.
Analysis presented in this section is generalization of our previous paper
\cite{Kluson:2019xuo} so that we recommend this paper for more details. As was argued previously in
\cite{Harmark:2017rpg,Kluson:2018egd,Harmark:2018cdl,Harmark:2019upf} non-relativistic string in torsional Newton-Cartan background can be naturally defined with the help of
 T-duality along light-like direction of relativistic string. Since as we argued in \cite{Kluson:2019qgj} such a transformation is problematic it is natural to consider an extended action with two auxiliary fields $\lambda^+,\lambda^-$
 \cite{Bergshoeff:2018yvt}. This extended action is chosen in such a way that when we solve equation of motion for $\lambda^+,\lambda^-$ we obtain original relativistic string with
light-like isometry. Explicitly, let us consider following Lagrangian density
\begin{eqnarray}\label{mLu}
& &\mL=-\frac{T}{2}N\sqrt{\omega}
[-\nabla_n x^M G_{MN}\nabla_n x^N+\frac{1}{\omega}
\partial_\sigma x^M\partial_\sigma x^NG_{MN}-
\nonumber \\
& &-2\nabla_n x^M
G_{M u}\nabla_n u+\frac{2}{\omega}\partial_\sigma x^MG_{M u}\partial_\sigma u
-\nabla_n u G_{uu}\nabla_n u+\frac{1}{\omega}
\partial_\sigma u G_{uu}\partial_\sigma u+\nonumber \\
& &+\lambda^+(\nabla_n u-\frac{1}{\sqrt{\omega}}\partial_\sigma u)\bY^++\lambda^-
(\nabla_n u+\frac{1}{\sqrt{\omega}}\partial_\sigma u)\bY^-
+\lambda^+\lambda^-]-\nonumber \\
& &-TB_{MN}
\partial_\tau x^M \partial_\sigma x^N-T\partial_\tau x^M B_{M u}\partial_\sigma u-
\partial_\tau u B_{u N}\partial_\sigma x^N \ , \nonumber \\
\nonumber \\
\end{eqnarray}
where  $\bY^+=\sqrt{G_{yy}} \ , \bY^-=-\sqrt{G_{yy}}$. In (\ref{mLu}) wee  used $1+1$ form
of the world-sheet metric where
\begin{equation}
\gamma_{\alpha\beta}=\left(\begin{array}{cc}
-N^2+N^\sigma \omega N^\sigma & N^\sigma \omega \\
N^\sigma \omega & \omega \\ \end{array}
 \right) \ , \quad \nabla_n=\frac{1}{N}(\partial_\tau-N^\sigma \partial_\sigma ) \ .
\end{equation}
We see from the Lagrangian density that there is exceptional direction labeled with $u$ since solving equations of motion for $\lambda^+,\lambda^-$ and inserting back to the Lagrangian density (\ref{mLu})
we obtain following contribution
\begin{eqnarray}
-[\nabla_n u \nabla_n u-\frac{1}{\omega}\partial_\sigma u
\partial_\sigma u]\bY^+\bY^-=[\nabla_n u \nabla_n u-\frac{1}{\omega}\partial_\sigma u
\partial_\sigma u]G_{uu}
\end{eqnarray}
so that terms proportional to $(\nabla_n u)^2 $ and
$(\partial_\sigma u)^2$ are zero and hence the Lagrangian density
effectively describes motion of the string in the background with
light-like isometry. However the advantage of the Lagrangian
(\ref{mLu}) is that it contains term quadratic in time derivative of
$u$ and hence we can easily perform canonical analysis. This has
been done in \cite{Kluson:2019qgj,Kluson:2019xuo} with the result
\begin{eqnarray}
& &H=\frac{N}{2\sqrt{\omega}T}
[ \pi_M G^{MN}\pi_N+2\pi_M G^{M u}
(\pi_u+\frac{T}{2}\tlambda^+\bY^++
\frac{T}{2}\tlambda^-\bY^-)+\nonumber \\
& &+(\pi_u+\frac{T}{2}\tlambda^+\bY^++
\frac{T}{2}\tlambda^-\bY^-)G^{uu}
(\pi_u+\frac{T}{2}\tlambda^+\bY^++
\frac{T}{2}\tlambda^-\bY^-)+\nonumber \\
& &+T^2\partial_\sigma x^MG_{MN}\partial_\sigma x^N
+2T^2\partial_\sigma x^M G_{M u}\partial_\sigma u+
T^2\partial_\sigma u G_{uu}\partial_\sigma u-\nonumber \\
& &-T^2\tlambda^+\partial_\sigma u \bY^++
T^2\tlambda^-\partial_\sigma u \bY^-+T^2\tlambda^+\tlambda^-]+
N^\sigma \mH_\sigma=N^\tau \mH_\tau+N^\sigma \mH_\sigma \ .
\nonumber \\
\end{eqnarray}
where we performed rescaling
\begin{equation}
\sqrt{\omega }\lambda^+=\tlambda^+ \ ,  \quad
\sqrt{\omega}\lambda^-=\tlambda^-
\end{equation}
and in the final step also  $N^\tau=\frac{N}{2\sqrt{\omega}T}$.
Further,
\begin{equation}
\pi_M=p_M+TB_{MN}\partial_\sigma x^N+TB_{Mu}\partial_\sigma u \ ,
\quad
\pi_u=p_u+TB_{uM}\partial_\sigma x^M
\end{equation}
It is convenient to introduce common notation $\hM=(M,u)$ so that the Hamiltonian
constraint has the form
\begin{eqnarray}
\mH_\tau=(\pi_{\hM}
+\frac{T}{2}\tlambda^+\bY^+_{\hM}+\frac{T}{2}\tlambda^-\bY^-_{\hM})
G^{\hM\hN}(\pi_{\hN}
+\frac{T}{2}\tlambda^+\bY^+_{\hN}+\frac{T}{2}\tlambda^-\bY^-_{\hN})+
\nonumber \\
+T^2\partial_\sigma \tx^{\hM}G_{\hM\hN}\partial_\sigma x^{\hN}-T^2\tlambda^+
\partial_\sigma \tx^{\hM}\bY_{\hM}^++T^2\tlambda^-\partial_\sigma \tx^{\hM}
\bY_{\hM}^-+T^2\tlambda^+\tlambda^- \  , \nonumber \\
\end{eqnarray}
where we also introduced $\bY^\pm_{\hM}=(\overbrace{0,\dots,0}^d,\bY^\pm)$.

We showed in \cite{Kluson:2019qgj,Kluson:2019xuo} that performing T-duality
 along $u-$direction we obtain
non-relativistic string in torsional Newton-Cartan geometry. We further analyzed
T-duality of non-relativistic string along spatial direction in \cite{Kluson:2019xuo} where
we were not able to study it in the most general case of non-zero NSNS two form field.
Now we fill this gap and consider  T-duality transformations along $k-$spatial directions that we label with $y^i$. Then T-dual coordinates are related to the original ones by following relations
\begin{equation}
p_i=-T\partial_\sigma \ty_i \ , \quad \tp^i=-T\partial_\sigma y^i \ , \quad
p_u=-T\partial_\sigma \eta \ , \quad \tp_\eta=-T\partial_\sigma u \ .
\end{equation}
In what follows we introduce common notation where
$\tx_m=(\ty_i,\eta), \tp^m=(\tp^i,\tp_\eta)$ and
$\bY^\pm_m=(\overbrace{0,\dots,0}^k,\bY^\pm)$. Further, remaining
coordinates are denoted as $x^\mu$ together with conjugate momenta
$p_\mu$. Then after T-duality transformation we get
\begin{eqnarray}
& &\pi_\mu=k_\mu-B_{\mu m}\tp^m \ , \nonumber \\
& & \pi_i=-T\partial_\sigma \ty_i-B_{im}\tp^m+TB_{i\mu}
\partial_\sigma x^\mu+\frac{T}{2}\tlambda^+\bY_i^++\frac{T}{2}\tlambda^-
\bY_i^-=-B_{im}\tp^m+T\bV_i \ , \nonumber \\
& & \pi_u=-T\partial_\sigma \eta-B_{um}\tp^m+TB_{u\mu}\partial_\sigma x^\mu
  +\frac{T}{2}\tlambda^+\bY_i^++\frac{T}{2}\tlambda^-
 \bY_i^-=-B_{um}\tp^m+T\bV_u
  \  \nonumber \\
\end{eqnarray}
or equivalently
\begin{equation}
\pi_n=T\bV_n-B_{nm}\tp^m \ , \bV_n=-\partial_\sigma \tx_n+B_{n\mu}\partial_\sigma x^\mu
+\frac{1}{2}\tlambda^+\bY_n^++\frac{1}{2}\tlambda^-
\bY_n^- \ .
\end{equation}
Then T-dual Hamiltonian constraint has the form
\begin{eqnarray}
& &\mH^T_\tau=k_\mu G^{\mu\nu}k_\nu+\tp^m \bH_{mn}\tp^n+
2\tp^m B_{m\hM}G^{\hM \mu}k_\mu+2Tk_\mu G^{\mu n}\bV_n
+T^2 \bV_m G^{mn}\bV_n+\nonumber \\
& &+2T\tp^m(B_{m\hM}G^{\hM n}\bV_n-G_{m\mu}\partial_\sigma x^\mu
+\frac{1}{2}\tlambda^+\bY_m^+-\frac{1}{2}\tlambda^-\bY_m^-)
+T^2
\partial_\sigma x^\mu G_{\mu\nu}\partial_\sigma x^\nu+
T^2\tlambda^+\tlambda^- \ ,
\nonumber \\
\end{eqnarray}
where
\begin{equation}
\bH_{mn}=G_{mn}-B_{m\hM}G^{\hM\hN}B_{\hN n} \ .
\end{equation}
Let us now proceed to the Lagrangian formulation of given theory.
Since the analysis is completely the same as in case of the
relativistic theory we immediately write the result
\begin{eqnarray}\label{mLnonT}
& &\mL=\tp^m\partial_\tau \tx_m \tp^m+p_\mu\partial_\tau x^\mu-H=
\nonumber \\
& &=\frac{1}{4N^\tau}[g'_{\tau\tau}-2N^\sigma g'_{\tau\sigma}+(N^\sigma)^2g'_{\sigma\sigma}]
-N^\tau T^2 g'_{\sigma\sigma}-Tb'_{\tau\sigma}-\nonumber \\
& &-\frac{T}{2}N^\tau\tlambda^+[\nabla_n \tx_m\bA^m-2T
\partial_\sigma \tx_m \bA^m+
\nabla_n x^\mu \bA_\mu
-2T\partial_\sigma x^\mu \bA_\mu ]-
\nonumber \\
& &-\frac{T}{2}N^\tau \tlambda^-[\nabla_n \tx_m \bB^m
+2T\partial_\sigma \tx_m \bB^m
+\nabla_n x^\mu \bB_\mu
+2T\partial_\sigma x^\mu \bB_\mu]
\nonumber \\
& &-N^\tau T^2 \tlambda^+\tlambda^- \bX \ ,
\nonumber \\
\end{eqnarray}
where
\begin{eqnarray}
& &\bA^m=\tG^{mn}(\bY_n^++B_{nk}\hG^{kl}\bY_l^+) \ , \quad
\bA_\mu=
(B_{\mu m}-G_{\mu m})\tG^{mn}(\bY_n^++B_{nk}\hG^{kl}\bY_l^+) \ ,
\nonumber \\
& &\bB^m=\tG^{mn}(-\bY_n^-+B_{nl}\hG^{lk}\bY_k^-) \ , \quad
\bB_\mu=(G_{\mu m}+B_{\mu m})\tG^{mn}(-\bY_n^-+B_{kl}\hG^{lk}\bY_n^-) \ ,
\nonumber \\
&  &\bX=\bY_m^+\tG^{mn}\bY_n^-+1 \ . \nonumber \\
\end{eqnarray}
The nature of the resulting T-dual string depends on the fact whether
$\bX$ vanishes or not. In case when $\bX\neq 0$ we can solve the equation of motion for $\tlambda^+,\tlambda^-$ as
\begin{eqnarray}
& &-\frac{1}{2}[\nabla_n \tx_m\bA^m-2T
\partial_\sigma \tx_m \bA^m+
\nabla_n x^\mu \bA_\mu
-2T\partial_\sigma x^\mu \bA_\mu ]-T\tlambda^-\bX=0 \ , \nonumber \\
& &-\frac{1}{2}[\nabla_n \tx_m \bB^m
+2T\partial_\sigma \tx_m \bB^m+\nabla_n x^\mu \bB_\mu
+2T\partial_\sigma x^\mu \bB_\mu]-T\tlambda^+\bX=0 \ .  \nonumber \\
\end{eqnarray}
These equations can be easily solved for $\tlambda^+$ and $\tlambda^-$. Then inserting these results into (\ref{mLnonT}) we obtain
\begin{equation}
\mL''=\frac{1}{4N^\tau}[g''_{\tau\tau}-2N^\sigma g''_{\tau\sigma}+(N^\sigma)^2
g''_{\sigma\sigma}]-N^\tau T^2 g''_{\sigma\sigma}-T b''_{\tau\sigma} \ ,
\end{equation}
where
\begin{eqnarray}
& &g''_{\alpha\beta}=G''^{mn}\partial_\alpha \tx_m\partial_\beta \tx_n+G''^{m}_{ \ \nu}
\partial_\alpha \tx_m\partial_\beta x^\nu+G''^{ \ n}_\mu\partial_\alpha x^\mu\partial_\beta \tx_n+G''_{\mu\nu}\partial_\alpha x^\mu \partial_\beta x^\nu \ ,
\nonumber \\
& &b''_{\tau\sigma}=B''^{mn}\partial_\tau \tx_m\partial_\sigma \tx_n
+B''^m_{ \ \nu}\partial_\tau \tx_m\partial_\sigma x^\nu+B''^{ \ n}_\mu
\partial_\tau x^\mu \partial_\sigma \tx_n+
B''_{\mu\nu}\partial_\tau x^\mu\partial_\sigma x^\nu \ ,
\nonumber \\
\end{eqnarray}
where primed components of the background metric and NSNS two form have the form
\begin{eqnarray}\label{Btwoprime}
&& G''^{mn}=G'^{mn}+\frac{1}{2\bX}(\bA^m\bB^m+\bA^n\bB^m) \ , \quad
G''^m_{ \ \mu}=G'^m_{ \ \mu}+\frac{1}{2\bX}(\bA^m\bB_\mu+\bB^m\bB_\mu) \ , \nonumber \\
& &G''_{\mu\nu}=G'_{\mu\nu}+\frac{1}{2\bX}(\bA_\mu\bB_\nu+\bA_\nu\bB_\mu) \ , \quad
B''^{mn}=B'^{mn}+\frac{1}{2\bX}(\bB^m\bA^n-\bB^n\bA^m) \ , \nonumber \\
& &B''^m_{ \ \mu}=B'^m_{ \ \mu}+\frac{1}{2\bX}(\bB^m\bA_\mu-\bA^m\bB_\mu) \ , \quad
B''_{\mu\nu}=B'_{\mu\nu}+\frac{1}{2\bX}(\bB_\mu \bA_\nu-\bB_\nu \bA_\mu) \ .
\nonumber \\
\end{eqnarray}
Clearly resulting string is relativistic string in the background fields given in (\ref{Btwoprime}).

As the second case let us consider the case when $\bX=0$. In this case the equation of motion for $\tlambda^\pm$ have the form
\begin{eqnarray}\label{eqXzero}
& &\nabla_n \tx_m\bA^m-2T
\partial_\sigma \tx_m \bA^m+
\nabla_n x^\mu \bA_\mu
-2T\partial_\sigma x^\mu \bA_\mu =0 \ , \nonumber \\
& &\nabla_n \tx_m \bB^m
+2T\partial_\sigma \tx_m \bB^m+\nabla_n x^\mu \bB_\mu
+2T\partial_\sigma x^\mu \bB_\mu=0 \ .  \nonumber \\
\end{eqnarray}
We multiply first equation with $\partial_\sigma \tx_m \bB^m+
\partial_\sigma x^\mu \bB_\mu$ and the second one with
$\partial_\sigma \tx_m\bB^m+\partial_\sigma x^\mu \bB_\mu$ and sum these
two equations so that we get
\begin{eqnarray}\label{Nsigma}
& &N^\sigma=\frac{\ba_{\tau\sigma}}{\ba_{\sigma\sigma}} \ ,  \quad
\ba_{\alpha\beta}=\frac{1}{2}\partial_\alpha \tx_m
(\bA^m\bB^n+\bA^n\bB^m)\partial_\beta\tx_n+\nonumber \\
& & +\frac{1}{2}\partial_\alpha \tx_m (\bA^m\bB_\mu+\bB^m\bA_\mu)
\partial_\beta x^\nu
+\frac{1}{2}\partial_\alpha x^\mu (\bA_\mu\bB^m+\bB_\mu\bA^m)
\partial_\beta \tx_m+
\frac{1}{2}\partial_\alpha x^\mu (\bA_\mu \bB_\nu+\bA_\nu\bB_\mu)
\partial_\beta x^\nu \ .
 \nonumber \\
\end{eqnarray}
On the other hand if we multiply two equations  in (\ref{eqXzero}) together and use
(\ref{Nsigma})
 we obtain $N^\tau$ to be equal to
\begin{equation}
N^\tau=\frac{\sqrt{-\det \ba_{\alpha\beta}}}{2T\ba_{\sigma\sigma}} \
 .
 \end{equation}
Finally we obtain T-dual string in the form of non-relativistic action
\begin{equation}
S=\int d\tau d\sigma \mL  \ ,
\end{equation}
where
\begin{equation}
\mL=-\frac{T}{2}\sqrt{-\det\ba}\ba^{\alpha\beta}g'_{\alpha\beta} -TB'_{\tau\sigma} \ ,
\end{equation}
where $\ba^{\alpha\beta}$ is matrix inverse to $\ba_{\alpha\beta}$.

Let us be more explicit and consider following
background with light-like isometry
\cite{Harmark:2017rpg,Kluson:2018egd,Harmark:2018cdl,Harmark:2019upf}
\begin{equation}
ds^2=g_{\hM\hN}dx^{\hM}dx^{\hN}=
2\tau (du-m)+h_{MN}dx^Mdx^N \ , \tau=\tau_M dx^M \  , \quad
m=m_M  dx^M  \ ,
\end{equation}
where NSNS two form has  following components
\begin{equation}
B_{MN} \ , B_{uM}=b_M \ .
\end{equation}
Let us now consider T-duality along singe spatial coordinate that we denote as $y$ together with T-duality along direction labeled by $u$. Then we have
\begin{eqnarray}
& &G_{\mu\nu}=h_{\mu\nu}-\tau_\mu m_\nu-m_\nu \tau_\mu\equiv \hh_{\mu\nu} \ , \nonumber \\
& &G_{uu} \ , \quad  G_{uy}=\tau_y \ , \quad G_{u\mu}=\tau_\mu \ ,  \nonumber \\
& &G_{yy}=h_{yy}-2\tau_y m_y \ , \quad G_{y\mu}=h_{y\mu}-\tau_ym_\mu-\tau_\mu m_y \
\nonumber \\
\end{eqnarray}
so that the metric $G_{mn}$ and two form $B_{mn}$ have  following form
\begin{equation}
G_{mn}=\left(\begin{array}{cc}
G_{uu} & \tau_y \\
\tau_y & h_{yy}-2\tau_y m_y \\ \end{array}
\right) \ , \quad  B_{mn}=\left(\begin{array}{cc}
0 & b_y \\
-b_y & 0 \\ \end{array}\right)
\end{equation}
so that metric inverse to $G_{mn}$ is equal to
\begin{equation}
\hG^{mn}=\frac{1}{G_{uu}(h_{yy}-2\tau_y m_y)-\tau_y^2}\left(\begin{array}{cc}
h_{yy}-2\tau_y m_y & -\tau_y \\
-\tau_y & G_{uu} \\ \end{array} \right) \ .
\end{equation}
We would like to consider situation when T-dual string is again non-relativistic string
that is ensured when $\bX=0$. Since $\bY^\pm_u=\pm \sqrt{G_{uu}}$ we
see that in order to ensure this condition we should have $\tG^{uu}=\frac{1}{G^{uu}}$.
In order to ensure this requirement we firstly impose condition that $\tau_y=0$ so that the matrices $\hG_{mn}$ and $\tG^{mn}$ have the form
\begin{equation}
\hG_{mn}=\left(\begin{array}{cc}
G_{uu}-\frac{b_y b_y}{G_{yy}} & 0 \\
0 & G_{yy}-\frac{b_y b_y}{G_{uu}} \\
\end{array}
\right) \ , \quad  \tG^{mn}=
\left(\begin{array}{cc}
\frac{1}{
G_{uu}-\frac{b_y b_y}{G_{yy}}} & 0 \\
0 & \frac{1}{G_{yy}-\frac{b_y b_y}{G_{uu}}} \\
\end{array}
\right)
\end{equation}
and hence $\bX=0$ on condition that
 $b_y=0$. Then
\begin{eqnarray}
& &\bA^u=\frac{1}{\sqrt{G_{uu}}} \ , \quad  \bA^y=0 \ , \quad
\bA_\mu=-(b_\mu+\tau_\mu)\frac{1}{\sqrt{G_{uu}}} \ ,
\nonumber \\
& &\bB^u=\frac{1}{\sqrt{G_{uu}}} \ , \quad \bB^y=0 \ , \quad
\bB_u=(\tau_\mu-b_\mu)\frac{1}{\sqrt{G_{uu}}} \
\nonumber \\
\end{eqnarray}
and hence we get
\begin{eqnarray}
\ba_{\alpha\beta}
=\frac{1}{G_{uu}}((\partial_\alpha \eta-b_\alpha)(\partial_\beta\eta-b_\beta)-\tau_
\alpha\tau_\beta) \ .  \nonumber \\
\end{eqnarray}
Finally we obtain transformation rules for components of the metric
\begin{eqnarray}
& &G'_{\mu\nu}=\hh_{\mu\nu}-\frac{1}{G_{uu}}\tau_\mu \tau_\nu
-\frac{\hh_{\mu y}\hh_{y\nu}}{G_{yy}}+\frac{1}{G_{uu}}b_\mu b_\nu
-\frac{1}{G_{yy}}B_{\mu y}B_{y\nu}=\nonumber \\
& &=\hh'_{\mu\nu}+\frac{1}{G_{uu}}
(b_\mu b_\nu-\tau_\mu \tau_\nu) \ , \quad
G'_{\eta\eta}=\frac{1}{G_{uu}} \ ,  G'_{\ty\ty}=\frac{1}{G_{yy}} \ ,
\nonumber \\
& &G'_{\eta\nu}=-\frac{b_\nu}{G_{uu}}=G'_{\nu\eta} \ , \quad  G_{\ty \mu}=-\frac{1}{G_{yy}}B_{y\mu}=G_{\mu\ty} \ , \nonumber \\
& &B'_{\mu\nu}=B_{\mu\nu}+\frac{1}{G_{uu}}(\tau_\mu b_\nu-b_\mu\tau_\nu)+\frac{1}{G_{yy}}
(\hh_{\mu y}B_{y\nu}-B_{\mu y}\hh_{y\nu})  \ , \nonumber \\
& &B'_{\mu \ty}=\frac{G_{\mu y}}{G_{yy}} \ , \quad
 B'_{\mu \eta}=\frac{\tau_\mu}{G_{uu}} \ .
\nonumber \\
    \end{eqnarray}
These are generalization of the T-duality rules that were determined in our previous paper
\cite{Kluson:2019xuo} to the case of non-zero NSNS two form. We also see that T-duality of non-relativistic string is again non-relativistic string  when we impose restriction
on the background metric given by conditions $\tau_y=b_y=0$ which agrees with the condition found in \cite{Kluson:2019xuo}.

\end{document}